\documentclass[12pt,preprint]{aastex}
\begin{document}
\title{Diffuse Ionized Gas inside the dwarf irregular galaxy NGC 6822\thanks{Based on observations collected at the European Southern Observatory, Chile, proposal number 69.C-0203(A)}  }
\author{A.M. Hidalgo-G\'amez}
\affil{Instituto de Astronom\'\i a, Universidad Nacional Aut\'onoma de M\'exico, Ciudad Universitaria, Aptdo. 70 264,
C.P. 04510, Mexico City, Mexico\\ and\\ Escuela Superior de F\'{\i}sica y Matem\'aticas, IPN, U.P. Adolfo L\'opez Mateos, Mexico City, Mexico}
\author{ and A. Peimbert} 
\affil{Instituto de Astronom\'\i a, Universidad Nacional Aut\'onoma de M\'exico, Ciudad Universitaria, Aptdo. 70 264,
C.P. 04510, Mexico City, Mexico}

\begin{abstract}

We have studied the differences between the diffuse ionized gas (DIG)  and the H\,{\sc ii} regions along a slit position in the local dwarf  irregular galaxy NGC 6822. The slit position passes through the two most prominent H\,{\sc ii} regions: Hubble V and Hubble X. Important differences have been found in the excitation, ionization, and [N\,{\sc ii}]$\lambda$6584\AA/H$\alpha$ and [S\,{\sc ii}]$\lambda$6717\AA/H$\alpha$ line ratios between the DIG and the H\,{\sc ii} locations. Moreover, the values of all the line ratios are not similar to those in the DIG locations of spiral galaxies but are very similar to the values in other irregular galaxies, such as IC 10. We also determined the rate of recombination using the He\,{\sc i} $\lambda$ 5875  line. Finally, we obtained a picture of the ionization sources of the DIG. We consider that the leakage of photons from the H\,{\sc ii} regions might explain most of the line ratios, except [N\,{\sc ii}]/H$\alpha$, which might be explained by turbulence. 


\end{abstract}
\section{Introduction}

The study of the diffuse ionized gas  (DIG) in irregular galaxies is very difficult, mainly due to the difficulties in distinguishing between proper H\,{\sc ii} regions and other regions of ionized gas with very low density. In spiral galaxies the diffuse gas is very easy to detect, as it is located either above the disc or in the interarm regions, neither of which exists in irregular galaxies. Consequently, such studies have been carried out for very few irregular galaxies. Hunter \& Gallagher (1994) solved this problem by studying the emission at the edges of a few irregular galaxies, while Martin (1997) in her sample did not put hard limits on the line ratios of either group. So far, all the places with ionized gas have been identified as H\,{\sc ii} regions, especially when only H$\alpha$ images were used (e.g. Hodge et al. 1989). Therefore, DIG locations in an H$\alpha$ image might be mistaken for low-excitation H\,{\sc ii} regions. 

In a previous study of the DIG in irregular galaxies (Hidalgo-G\'amez 2005, 2006), a distinction was made between DIG and H\,{\sc ii} regions. Such a distintion is important due to the different spectral characteristics of the H\,{\sc ii} and DIG locations: DIG exhibits lower excitation and larger [N\,{\sc ii}]$\lambda$6584\AA/H$\alpha$ and 
[S\,{\sc ii}]$\lambda$6717\AA/H$\alpha$ ratios. Such differences were first observed in spiral galaxies (e.g. Rand 1998; T\"ullman \& Dettmar 2000) and then  confirmed in irregular galaxies (Hidalgo-G\'amez 2005, 2006). Hidalgo-G\'amez \& Ram\'{\i}rez-Fuentes (2006) have shown how the inclusion of DIG locations as  part of an H\,{\sc ii} region has a drastic influence on the metallicity of the region. Moreover, when a proper comparison is made between the DIG in spiral and in irregular galaxies, it becomes clear that they are very different. This difference could be due to differences in the metallicity and the electronic temperature, as well as in the dust content and gas mass in both types of galaxies. However, the mechanisms involved in the ionization of this low-density gas might also be different. So it is very important to disentangle real H\,{\sc ii} regions from DIG in order to obtain reliable conclusions. 

At this stage of the investigation of the DIG properties in irregular 
galaxies, we are interested in comparing the line ratios found
in irregular galaxies to those in spirals. Also, we verify whether the 
present models, obtained specifically for the Milky Way and other high-metallicity spirals, can also explain the spectral lines in irregular 
galaxies. If there are any differences that are related to global 
properties of the interstellar medium (ISM), such as the metal abundance or the star formation rate 
(SFR), they could be of the utmost importance in the evolution of galaxies. Such 
relationships are studied elsewhere (Hidalgo-G\'amez 2006).     

In the present investigation we study the DIG inside NGC 6822. This is a 
close galaxy (d$\approx$ 500 kpc; Karachentsev 2005), defined by de 
Vaucouleurs et al. (1991) as a barred irregular with an oxygen abundance 
slightly lower than that of the Large Magellanic Cloud (LMC; Hidalgo-G\'amez et al. 2001, hereafter HGOM01; 
Peimbert et al. 2005). It is a small galaxy, with $r_{25} = 1.08$ kpc (Hidalgo-G\'amez \& Olofsson 1998) but an extension on the sky of 13.5'$\times$15.5'. Hodge et al. (1989) found a total of 145 H\,{\sc ii} regions, some of which are extremely faint. The two brightest H\,{\sc ii} regions, -Hubble V and Hubble X, located in the northern part of the galaxy-, and the DIG in their vicinities are the targets here. The main reason is that they are very close to each other, separated only by $300$ pc; both have very similar metal content and ionizing stellar temperature (e.g. HGOM01). Therefore, the OB stellar populations should be very similar in both of them, as should the ionizing mechanisms of the regions. In addition, and contrary to what happens in IC 10, the SFR is low, of order of 0.04 $M\odot$ yr$^{-1}$ (Wyder 2001), slightly lower than the typical values in irregular galaxies (Hunter \& Elmegreen 2004);  the existence of giant bubbles has not been reported there, although de Blok \& Walter (2000) expressed that the H\,{\sc i} is very disturbed. This galaxy, a very typical irregular galaxy, is in a sense an ideal candidate for comparing the ionization mechanics of the DIG among irregular galaxies and between irregulars and spirals. 

The paper is structured as follows.: The information on the acquisition of the data,  the reduction procedure, and the analysis of the data are given in Section 2. Section 3 describes the main characteristics of the DIG in the region. The ionization source is studied in Section 4, while a discussion on the possible geometry is given in Section 5. Finally, conclusions are outlined in Section 6.

\section{Observations and data reduction}

\begin{figure*}
   \centering
   \includegraphics[width=18cm]{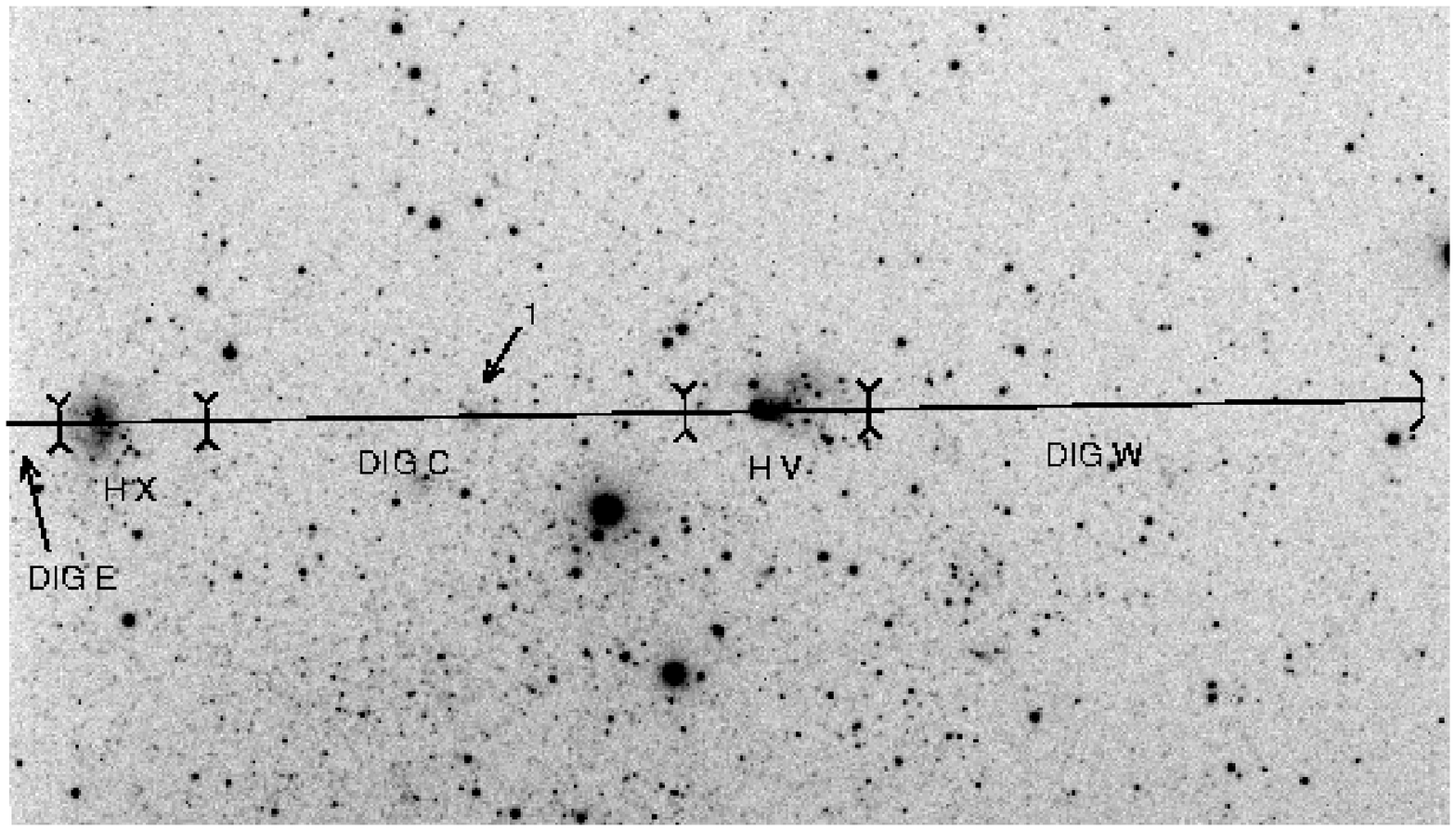}
      \caption{ VLT image of the northern part of NGC 6822. The image of 392" x 224" is centered at alpha=19$^h$ 44$^m$ 53.4$^s$,  delta=-14$^o$ 43' 15".
The total area of the five regions adds up to 400" x 0.51"(see text).
Most of the extraction aperture is shown in the figure; however
16'' of the DIG E region are not in the figure.
Point ``1'' indicates a region of low hydrogen emission (see text). }
         \label{fig1}
   \end{figure*}

The data were long-slit observations acquired with the Focal Reducer/Low
Dispersion Spectrograph, (FORS1), at the  Melipal Very Large Telescope (VLT) in Cerro Paranal
(Chile). The detector is a 2048$\times$2048 pixel CCD; it was used with an image
scale of 0.2"/pixel providing a slit length of 5.8'. The slit width was set to
0.51" and the slit orientation was almost east-west (position angle 91$^o$) in order
to observe Hubble V and Hubble X simultaneously. Figure ~\ref{fig1} shows the
region under study, with the slit superimposed. Two grism settings were used in
these observations: GRIS600B+12, which provided a spectral coverage in
the 3450-5900 {\AA} range, and  GRIS600R+14 with the filter
GG435, which provided a spectral coverage in the 5250-7450 {\AA}
range. The resolution for the emission lines observed with the blue grism is
given by $\Delta \lambda \sim \lambda /1300$, while the resolution with the red grism is given by
$\Delta \lambda \sim \lambda /1700$. There were three integrations, 12 minutes
each, in the blue setting, and three integrations, 10 minutes
each, in the red setting. The spectra were reduced following the standard procedure using the IRAF reduction package. Bias and sky twilight flat-fields were used for the calibration of the CCD response.  Ne-Ar-He lamps were used for the wavelength calibration. The spectra were corrected for atmospheric extinction using the La Silla extinction law. Four standard stars were observed in order to perform the flux calibration. More details on the reduction procedure can be found in Peimbert et al. (2005).

These two spectra were divided into a total of $40$ one-dimensional spectra,
each being the average of 50 rows corresponding to 10$''$ on the
sky. Although the seeing conditions were good enough, the low signa-to-noise (S/N) at very low
surface brightness due to the small width of the slit, produced a very large
number of pixels. For each of these spectra, the intensity of the recombination
lines H$\alpha$ and H$\beta$, as well as the forbidden lines
[O\,{\sc ii}] $\lambda$3727, [O\,{\sc iii}] $\lambda$5007, [N\,{\sc ii}] $\lambda$6583,
[S\,{\sc ii}] $\lambda$6717, and [S\,{\sc ii}] $\lambda$6731, were measured. In addition, we
obtained a total of five spectra (hereafter, referred as the integrated spectra), covering
the two main H\,{\sc ii} regions and the three DIG regions: two at the edge of
the slit (DIG E and DIGW) and the third one in between the two H\,{\sc ii}
regions (DIG C). The lengths of these integrated spectra were as follows: DIG E=30'',
Hubble X=40'', DIG C=130'' Hubble V=50'' and DIG W=150''. For these five spectra, the helium recombination
line at 5876\AA~ was also measured, in addition to the previously
mentioned lines.

The lines in both sets of spectra were normalized to H$\beta$, absorption-corrected following McCall et al. (1985) and extinction-corrected using 

 $$ C_{\beta} = -{1 \over f(\lambda) - f(H\beta)} ~\ln \lbrack{I(\lambda_o) 
\slash I(H\beta_o)\over I(\lambda) \slash I(H\beta)}\rbrack$$ 
where $I(\lambda_o)\slash I(H\beta_o)$ is the observed intensity of the line 
relative to H$\beta$, $I(\lambda)\slash I(H\beta)$ is the theoretical value 
determined for an optically thick gas (Brocklehurst 1971), and $f(\lambda)$ 
is the adopted extinction law; in this paper we used the Whitford modified 
law (Savage \& Mathis 1979). One caveat in the extinction correction is the H$\beta$ absorption in the stellar continuum present at most DIG locations. This underlying absorption can  significantly change the H$\beta$ emission intensity.  Therefore, the extinction might be somewhat 
artificial, and $C_{\beta}$ can produce odd values of the line intensities, especially [O\,{\sc ii}] $\lambda$3727. As can be seen, the values of 
these coefficients inside the H\,{\sc ii} regions do not match the previous 
ones (HGOM01; Peimbert et al. 2005). In this investigation the extinction-corrected intensities are be preferred. Those ratios for which the differences between the reddened and non-reddened intensities are important (up to 30\%), such as [O\,{\sc ii}]/H$\beta$, [O\,{\sc ii}]/H$\alpha$, and [O\,{\sc iii}]/H$\alpha$, are not considered.

After these corrections, the  line ratios [O\,{\sc iii}] $\lambda$5007/H$\beta$, 
[N\,{\sc ii}] $\lambda$6584/H$\alpha$, [S\,{\sc ii}] $\lambda$6717/H$\alpha$, [N\,{\sc ii}] $\lambda$6584/[S\,{\sc ii}] $\lambda$6717, and [S\,{\sc ii}] $\lambda$6717/[S\,{\sc ii}] $\lambda$6731 were determined for both, the one-dimensional and the integrated spectra sets. In the latter, the He\,{\sc i} $\lambda$5876/H$\alpha$ ratio was also determined. As this helium line blends with a sky line, the values tabulated here might be slightly larger than the real ones. These values are given in Table 1 for the integrated spectra, with their uncertainties. The formal errors were determined from $$ \sigma_t = (\sigma_c^2 +  \sigma_r^2 + \sigma_a^2)^{1/2}$$ with $\sigma_c$ representing the uncertainties at the level of the spectral continuum with respect to the line;  $\sigma_r$ are those uncertainties  introduced by the reduction procedure, and $\sigma_a$ are the uncertainties due to the extinction and absorption corrections. They are of  order  less than $5$\%. Another $10$\% was added to the uncertainties of the line ratios based on the differences between the low-density limit, [S\,{\sc ii}] $\lambda$6717/[S\,{\sc ii}] $\lambda$6731 = $1.44$  (Osterbrock 1989), and the density at the DIG locations. The variations in the continum in 
the vicinity of the helium line increase the uncertainties up to $50 \%$, but the line is clearly detected.

\section{Diffuse Ionized Gas inside NGC 6822}

The best parameter for defining the DIG is the density. But density is also one of the most difficult parameters to determine from 
long-slit spectroscopy data, especially in the low-density regime. The main problem is that in this regime, the 
differences in the [S\,{\sc ii}] $\lambda$6717/[S\,{\sc ii}] $\lambda$6731 ratio between the 
H\,{\sc ii} regions, with typical densities of $100$ cm$^{-3}$, and the DIG 
with values of $10$ cm$^{-3}$, are of order $0.1$ (see Fig. 5.3 in Osterbrock 
1989), which normally is smaller than the uncertainties in the ratio. Therefore, another parameter has to be used for the definition of DIG.

\begin{figure}
   \centering
   \includegraphics[width=15cm]{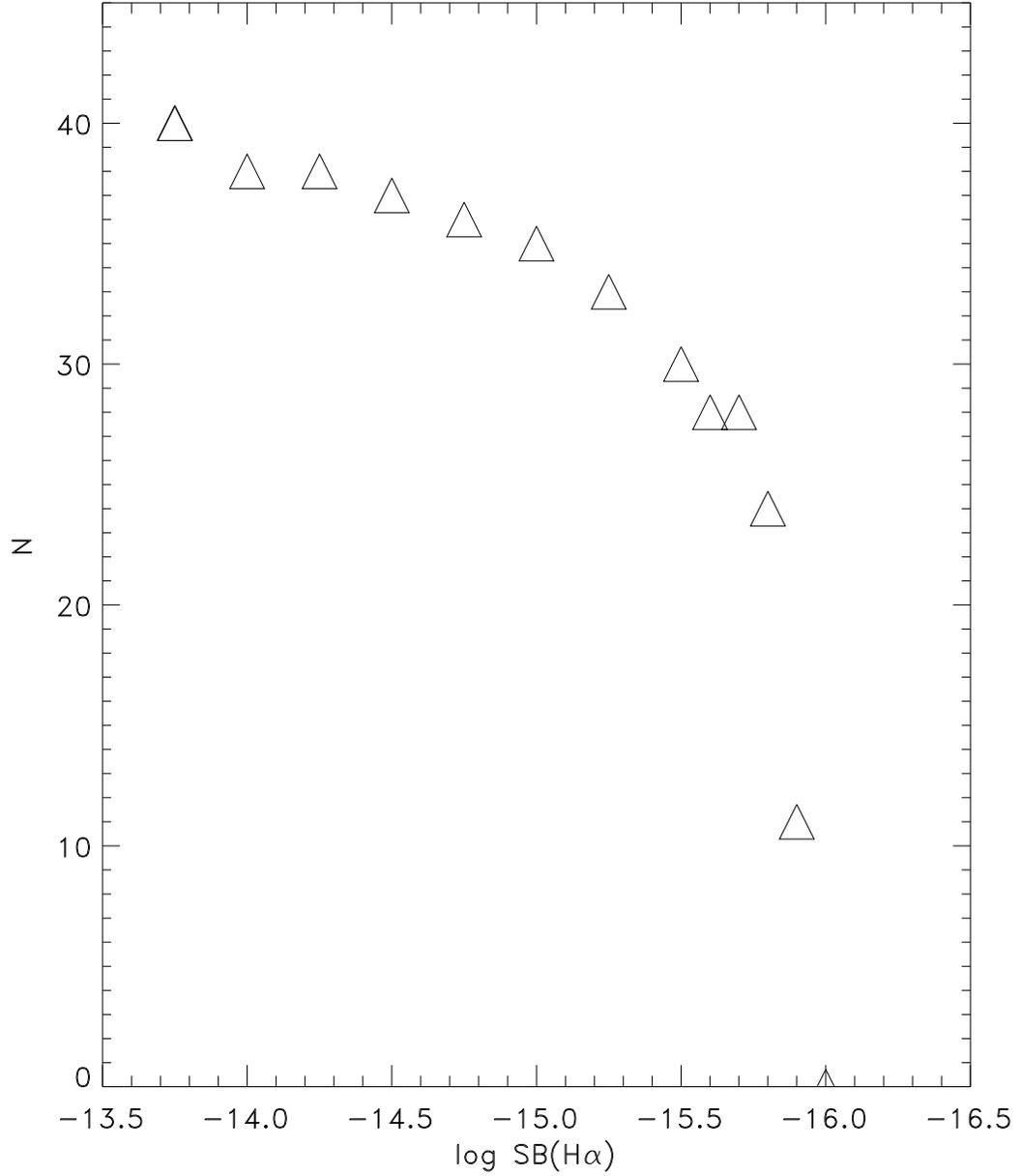}
      \caption{  Cumulative function of the surface brightness in H$\alpha$ for all the data along the slit in NGC 6822. The turnoff point represents the limiting surface brightness value between DIG and H\,{\sc ii} regions. For this galaxy, it is -15.3.  }
         \label{fig2}
   \end{figure}

The most important parameter after the density is the emission measure ($EM$) (Greenawalt et al. 1998) related to the surface brightness in H$\alpha$ by the expression $$EM = (7.22 \times 10^{13}) SB(H\alpha) T^{0.96}$$, where $T$ is the electronic 
temperature and $SB(H\alpha)$ is the surface brightness in H$\alpha$ in ergs cm $^{-2}$ s$^{-1}$ arcsec$^{-2}$ (M. Richer 2001, private communication). This equation is equivalent to other formulations existing in the literature (Reynolds et al. 1999). In spiral galaxies, values of EM between $2$  and $100$ cm$^{-6}$ pc have been considered (Greenawalt et al. 1998; 
Reynolds 1984). In  previous studies of the DIG, values of $EM = 2$ cm$^{-6}$ 
pc 
for IC 10 DIG locations (Hidalgo-G\'amez 2005) and $EM = 8$ cm$^{-6}$ pc for GR 8 and 
ESO 245-G05 DIG locations (Hidalgo-G\'amez 2006) were determined using the cumulative function of the surface 
brightness in H$\alpha$, with Galactic extinction corrected. A similar 
approach has been taken here. On Figure ~\ref{fig2} this function with the  data of NGC 6822 is shown. Two straight lines can be fitted to this curve. The intersecting point corresponds to the surface brightness [hereafter referred to as SB(H$\alpha$)], which divides the sample into
 H\,{\sc ii} regions and DIG. In Figure ~\ref{fig2} this point is located at $5 \times 10^{-16}$ 
ergs cm$^{-2}$ s$^{-1}$ arcsec$^{-2}$. This surface brightness gives an $EM  \approx 200$ cm$^{-6}$pc for $10,000$ K (see Table 1). This value is much larger than the values obtained for the galaxies studied so far for 
similar values of $T_e$, and only similar to the value in DDO 50 (A.M. Hidalgo-G\'amez 2007, in preparation).  
 
Using this surface brightness, we have divided the $40$ spectra into two groups: $31$
are DIG locations and $9$ are H\,{\sc ii} regions. It has to be realized that because of the large size of the 
spectra ($50$ rows),  H\,{\sc ii} regions may include a part of the DIG and vice versa. The H\,{\sc ii} regions of $\approx$ $150$ and 
$125$ 
pc (projected sizes), correspond to Hubble V and Hubble X, respectively. 
These sizes are comparable to the sizes of the regions obtained in 
previous studies (HGOM01). 

In Table 1, the line ratios measured from the integrated spectra are listed for the three DIG and the H\,{\sc ii} locations. Along with them, the extinction parameter as described previously, the S/N in the H$\alpha$ line, and the 
electronic temperature of the DIG are shown. 
The electronic temperature in the DIG was obtained using both oxygen and nitrogen 
lines. For the oxygen temperature we assumed an 
abundance of $8.34$ (Peimbert et al. 2005) and T([O\,{\sc ii}]) = T([O\,{\sc iii}]) and 
searched 
for the temperature that would reproduce the intensities of [O\,{\sc ii}]$\lambda$3727. 
For the nitrogen temperature we assumed an abundance of $7.05$ (Peimbert et al. 
2005) and a nitrogen degree of ionization equal to that of oxygen.

\subsection{DIG along the slit position}

\begin{figure}
   \centering
   \includegraphics[width=15cm]{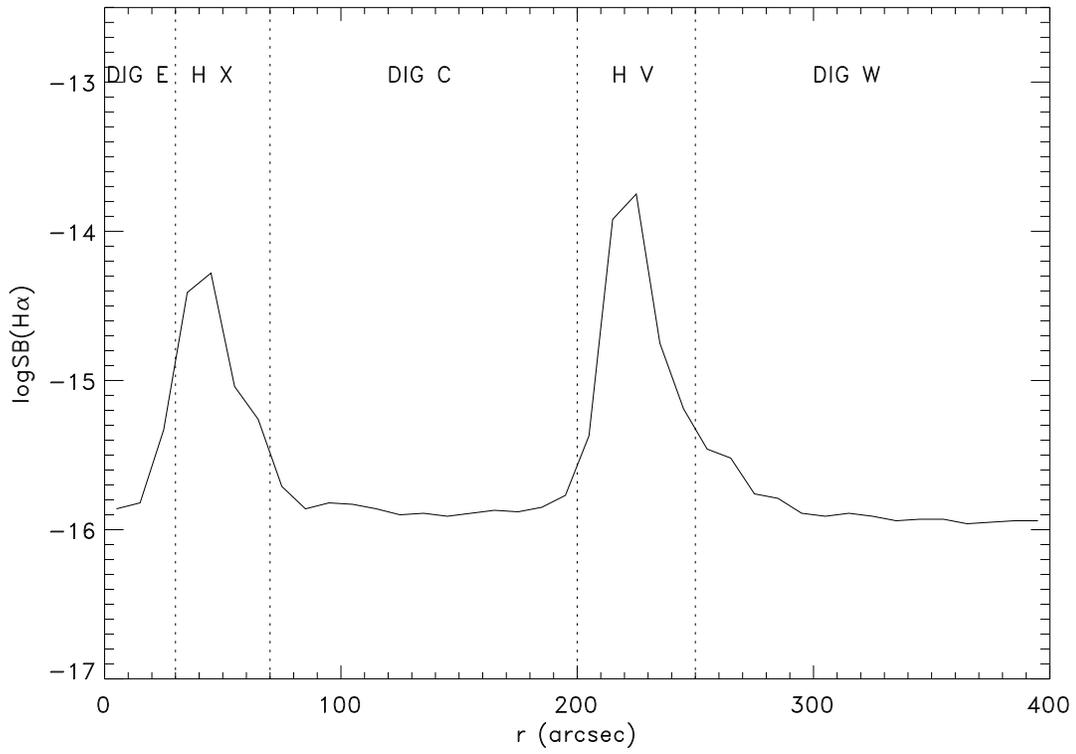}
 \caption{ SB(H$\alpha$) along the slit position 
in NGC 6822. The dotted lines include the H\,{\sc ii} regions. East  
is to the left. }
         \label{fig3}
\end{figure}

\begin{figure}
   \centering
   \includegraphics[width=15cm]{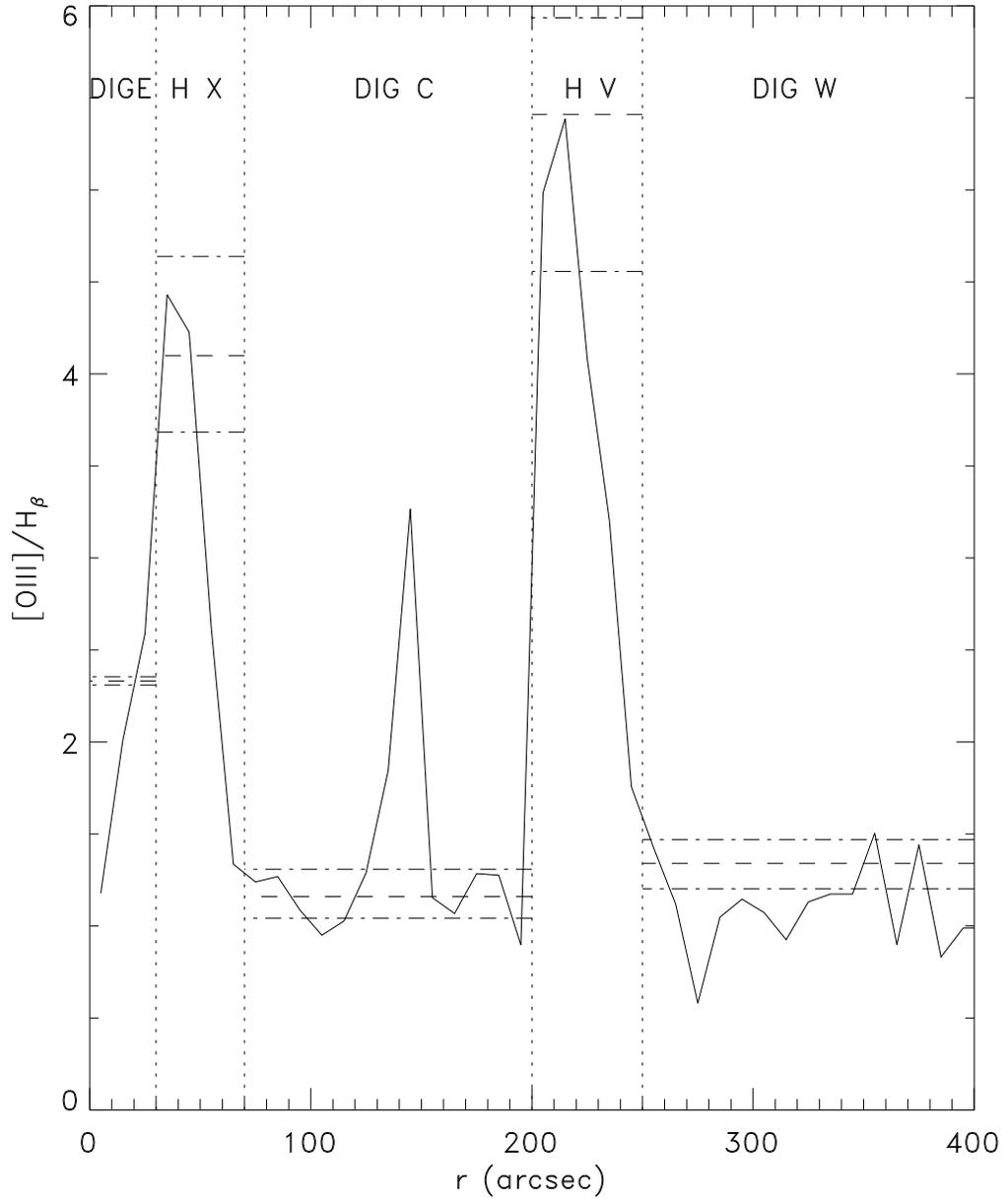}
 \caption{ Excitation, defined as [O\,{\sc iii}]/H$\beta$, along the slit. The dotted lines include the H\,{\sc ii} regions, while the dashed lines  correspond to the integrated spectra values whith error bars ($dot-dashed$ lines). Orientation is as in Fig. ~\ref{fig3}. }
         \label{fig4}
   \end{figure}

\begin{figure}
   \centering
   \includegraphics[width=15cm]{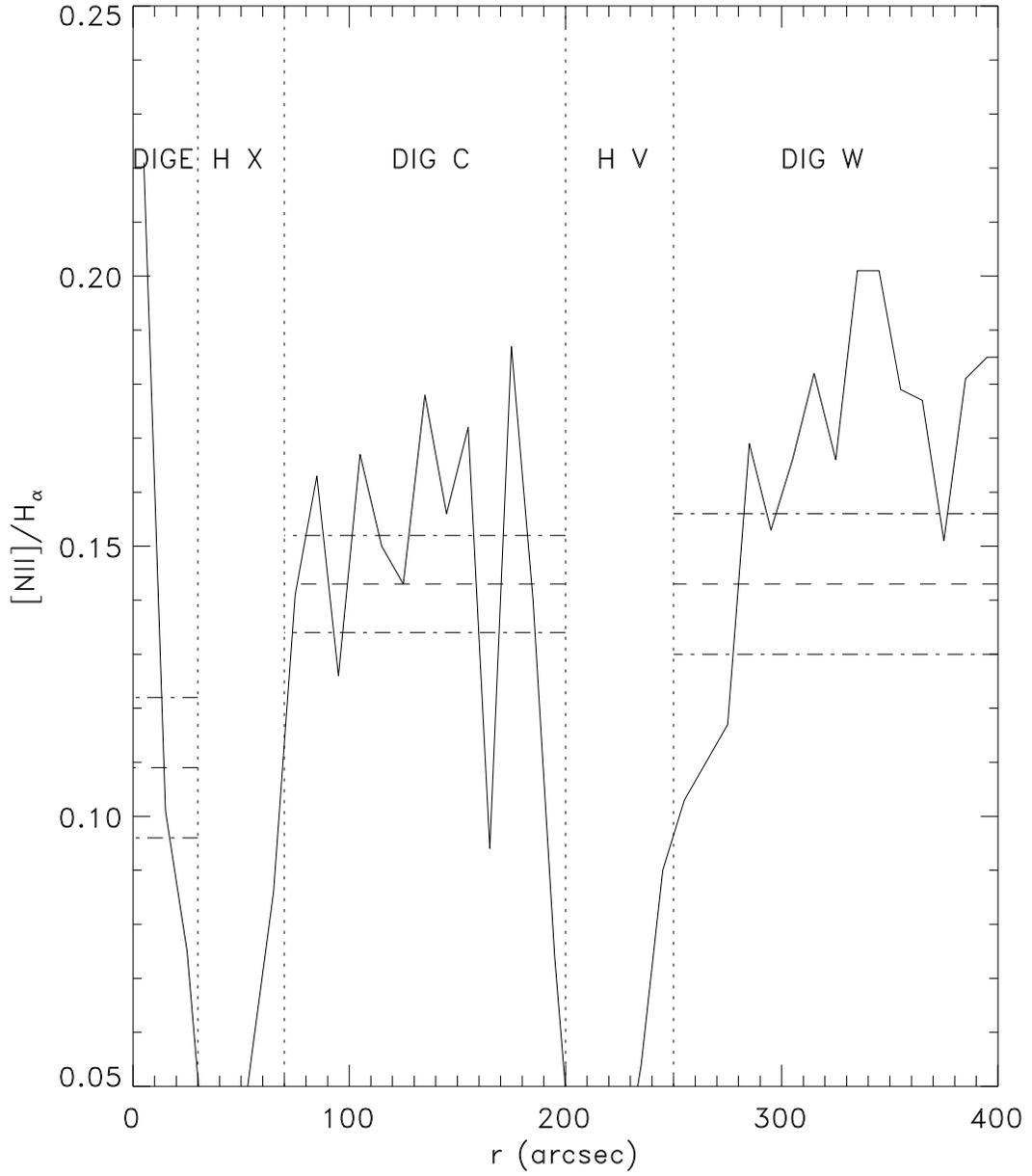}
      \caption{  [N\,{\sc ii}] $\lambda$6584/H$\alpha$ ratio along the slit. Symbols and orientation are as in Fig. ~\ref{fig4}.    }
         \label{fig5}
   \end{figure}

\begin{figure}
  \centering
   \includegraphics[width=15cm]{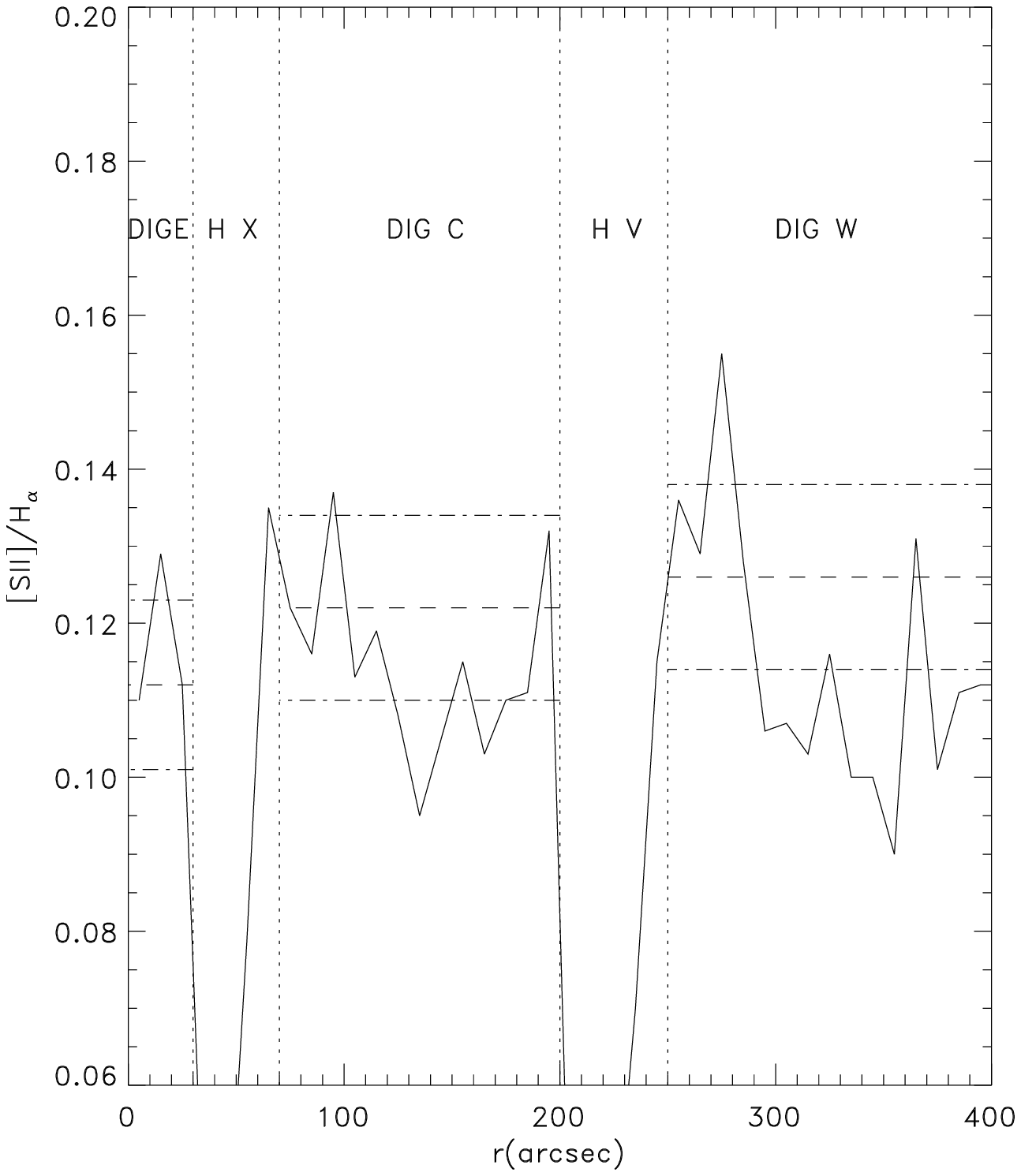}
      \caption{   [S\,{\sc ii}] $\lambda$6717/H$\alpha$ 
 ratio along the slit. Symbols and orientation are as in Fig. ~\ref{fig4}. }
         \label{fig6}
   \end{figure}

Figures ~\ref{fig3}-\ref{fig6} present values along the slit of the 
SB(H$\alpha$), the excitation, the ionization, [N\,{\sc ii}] $\lambda$6584/H$\alpha$, and [S\,{\sc ii}] $\lambda$6717/H$\alpha$. All the plots follow 
a similar structure: in the $x$-axis the spatial extension of the region under study  is 
plotted from the east to the west of the galaxy. The line ratios under consideration 
are shown at the $y$-axis. The dotted lines contain the H\,{\sc ii} regions. In all 
except Figure ~\ref{fig3}  the values of the integrated spectra are 
shown 
($dashed line$), and the dot-dashed lines correspond to their uncertainties.

Figure ~\ref{fig3} gives the log of the surface brightness in the 
H$\alpha$ line. The most interesting feature is the relatively constant SB for the DIG locations. The H\,{\sc ii} regions are clearly defined by their higher SB.

Figure  ~\ref{fig4} shows the excitation, defined as [O\,{\sc iii}]$\lambda$5007/H$\beta$, along the slit. Hubble V, located between $200''$ and $250''$ , has a larger excitation than Hubble X (between $30''$ and $70''$), which is in agreement with previous investigations (HGOM01). For both the DIG and H\,{\sc ii} regions the excitation from the integrated spectra agrees with the values from the one-dimensional spectra when error bars are taken into account. The typical error bar of the latter spectra in both H\,{\sc ii} regions is $11 \%$. For the DIG locations error bars are larger: as high as $21 \%$. The error bars are not shown in the figures for the sake of clarity.  Considering the 
DIG locations, their  excitation is higher than the results obtained from spiral galaxies (e.g. M31; Greenawalt et al. 1998). Actually, only eight locations have values of [O\,{\sc iii}]/H$\beta$ smaller than $1$, and only in one location is it smaller than $0.6$, at $275''$. Although this excitation is very high, other irregular galaxies have similar (DDO 53; Hidalgo-G\'amez \& Flores-Fajardo 2007) or even larger (IC 10; Hidalgo-G\'amez 2005) values. 

Concerning the DIG integrated spectra, the larger values of the excitation correspond to the region named DIG E. This region was averaged over only $125$ rows and therefore has a S/N lower than the other two DIG regions. Therefore, and considering the similar value of the excitation between regions W and C, the excitation might be overestimated in region E. 

Another important feature is the large peak between $130''$ and $150''$. It cannot be due to an excitation increment at these locations because the intensity of the [O\,{\sc iii}] $\lambda$5007 line is approximately constant. On the contrary, the intensity of H$\beta$ drops by a factor of $4$ and $3$, respectively, in these two spectra. The best explanation could be a severe absorption in the H$\beta$ intensity. Actually, a nebular condensation is clearly seen in Figure ~\ref{fig1} in this place (marked as $''1``$), which might be a young stellar cluster. A value of $1.20$ is obtained for the integrated spectra of DIG C when  this region is  excluded. This is consistent within the error bars with the value listed in Table 1. By comparing Figures ~\ref{fig3} and ~\ref{fig4}, it is clearly seen that those locations with high SB(H$\alpha$) are also those with the 
larger excitation, which actually correspond to the H\,{\sc ii} regions.

The ratios [N\,{\sc ii}] $\lambda$6584/H$\alpha$ and [S\,{\sc ii}] $\lambda$6717/H$\alpha$ are very interesting because DIG in spiral galaxies has very high values for them (Otte \& Dettmar 1999). Those ratios are shown in Figures ~\ref{fig5} and ~\ref{fig6}, respectively. In both cases, there is  a large increment of these ratios at the DIG locations. The values of the H\,{\sc ii} regions are very small, but in perfect agreement with more detailed investigations of these two regions (HGOM01).  Some of the one-dimensional spectra show large differences compared to the integrated spectra in [NII]/H$\alpha$. Examples include the spectrum at $330''$-$340''$ with [NII]/H$\alpha$ of $0.2 \pm 0.1$, and the spectrum at $160''$ with [NII]/H$\alpha$ of $0.09 \pm 0.01$. The reason for such discrepancy is likely the low S/N of the one-dimensional spectra (Rola \& Pelat 1994). The [NII]/H$\alpha$ ratio from the integrated spectra is almost indentical for DIG C and W and the mayority of the one-dimensional spectra of DIG C and W agree well with the integrated spectra values, when error bars are considered. Such error bars could be as high as $30 \%$ in some of the one-dimensional spectra.

[S\,{\sc ii}]$\lambda$6717/H$\alpha$ is a ratio related to shocks (Dopita 1993). 
For very large values ($> 0.3$), shocks must be important contributors to the intensity 
of this ratio. Two things can be noted from Figure ~\ref{fig6}: first, 
the values are very small, the higher value being $0.15 \pm 0.09$; there are no spectra along the slit where this ratio is 
large enough for shocks to be important. Second, although the ratio is larger in DIG 
locations than inside the H\,{\sc ii} regions by a factor of $3$, in the former the 
ratio is very constant, considering the uncertainties.  
Without the reddening correction, the ratio is slightly larger (by $0.02$), 
especially at low SB(H$\alpha$) locations. As for the [N\,{\sc ii}]/H$\alpha$ ratio, there are some discordant one-dimensional spectra in the [S\,{\sc ii}]/H$\alpha$ ratio at $135''$, $275''$, and $355''$, with values of $0.095 \pm 0.03$, $0.15 \pm 0.04$, and $0.09 \pm 0.03$, respectively. When uncertainties in these spectra are considered ($\approx 30 \%$), they agree with the integrated spectra values. Again, the [SII]/H$\alpha$ ratios in DIG C and W are almost identical.    

The ratio [S\,{\sc ii}] $\lambda$6717/[S\,{\sc ii}] $\lambda$6731 related to the density (Osterbrock 1989) has also been studied. But because one-third of the data go beyond the low-density limit, this plot is not shown. The same occurs 
for the integrated spectra (see Table 1). Another important feature is the 
uniformity of this ratio along the slit, showing no differences between DIG and 
H\,{\sc ii} locations. Therefore, it can be concluded that every place along 
the slit has very low density (n$_e$ $<$ $200$ cm$^{-3}$). Actually, very low density 
values are found inside the H\,{\sc ii} regions from the [O\,{\sc ii}] and [Cl\,{\sc iii}] 
doublets (Peimbert et al. 2005). 

\section{On the search for the ionization source}

\subsection{Photoionization}

According to Str\"omgren (1939), H\,{\sc ii} regions could be radiation- or 
density-bounded. If all the energy of the ionizing photons is used to ionize 
the atoms, the H\,{\sc ii} region is referred to as radiation-bounded. On the contrary, 
when the photons at the edge of the region still have enough energy to ionize 
atoms, it is called a density-bounded region. 

Considering H\,{\sc ii} regions as radiation-bounded, Mathis (1986) and 
Domg\"orgen \& Mathis (1994) made some predictions on the DIG line ratios:  
large values for both [S\,{\sc ii}] $\lambda$6717+6731/H$\alpha$ and 
[N\,{\sc ii}] $\lambda$6584/H$\alpha$ ($> 0.3$), very large [O\,{\sc ii}]/H$\alpha$ ($> 1$) values, and  very small [O\,{\sc iii}]/H$\alpha$ ($< 0.1$) and He \,{\sc i}/H$\alpha$ ($< 0.03$) values. We 
can compare them with the values from Table 1. The [O\,{\sc ii}]/H$\alpha$ and [O\,{\sc iii}]/H$\alpha$ ratios are not  used due to the extinction problems previously 
discussed. None of the other three lines has the value predicted by these 
photoionization models. The main weakness is that the models were optimized 
for Milky Way metallicity, whereas NGC 6822 has a metallicity 
$\approx 1/30$ Z$_{\odot}$. Castellanos et al. (2004) and Elwert \& Dettmar 
(2005) used the photoionization code CLOUDY with different input parameters 
and lower metallicities in order to reproduce the line ratios observed. 
Both are successful in reproducing the rise of the [N\,{\sc ii}] versus [S\,{\sc ii}] plot, but 
none of them can fit the excitation. When Monte Carlo simulations with 
different ionizing spectra and metallicities are performed (Wood \& Mathis 2004), a similar situation takes place.

Another prediction of the photoionization models is the relatively constant [N\,{\sc ii}] $\lambda$6584/H$\alpha$ line ratio between the DIG and H\,{\sc ii} regions. From Figure ~\ref{fig5} it can be seen that it does not happen here; in fact, there 
are strong variations along the slit.  Such variations can be due to differences in the content of nitrogen or in the electronic temperature. Thus, for example, differences of $600$ K in the $T_e$ can explain the small value of the spectrum at $165''$. Such differences, although large, are in fact observed in the outer 
part of Hubble V and Hubble X (see Fig. 8 in HGOM01). Therefore, they cannot be 
ruled out as a source of inhomogeneity in the [N\,{\sc ii}] $\lambda$6584/H$\alpha$ ratio.

\begin{figure}
   \centering
   \includegraphics[width=15cm]{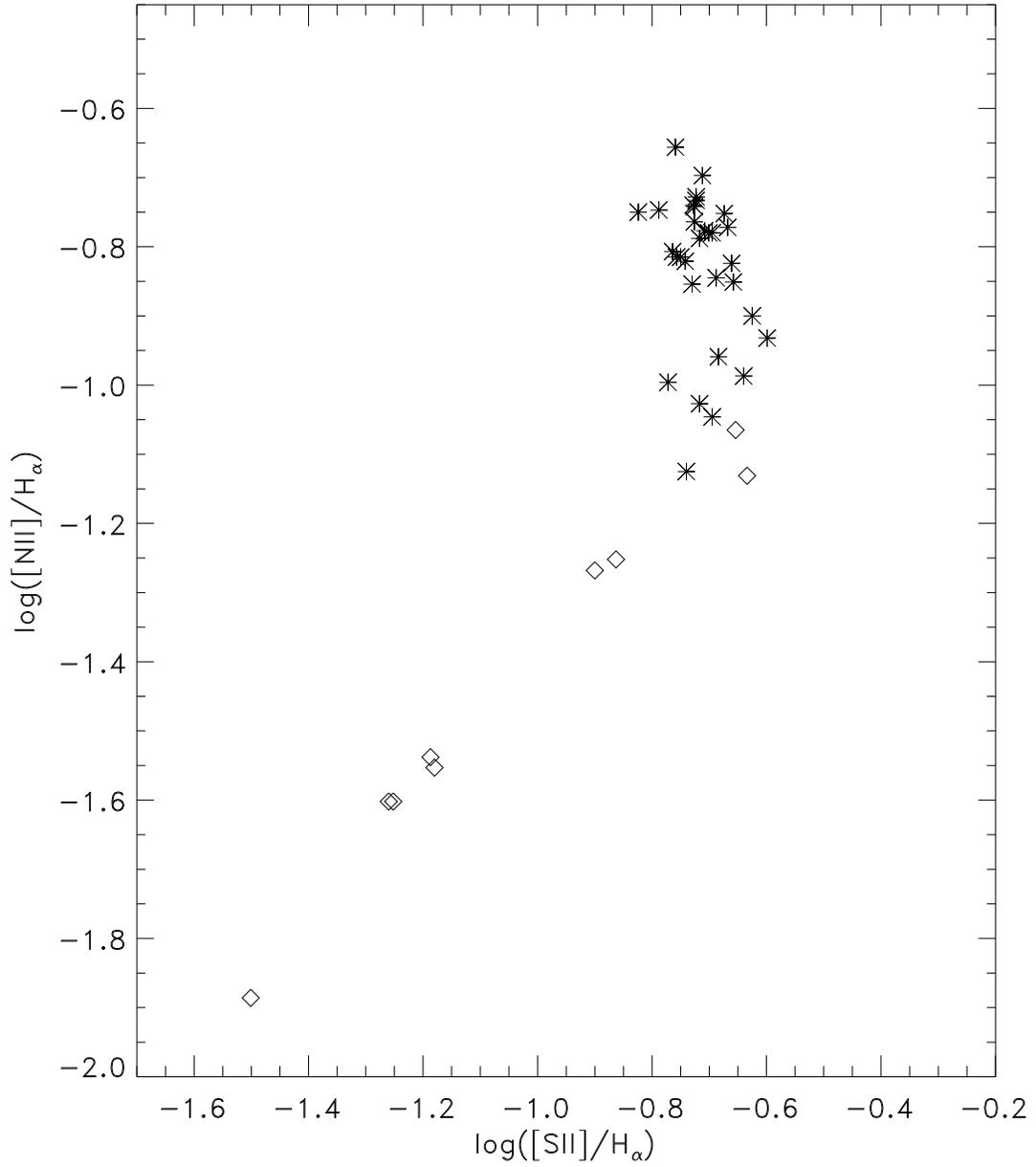}
      \caption{ log([N\,{\sc ii}]$\lambda$6584/H$\alpha$) vs. log 
([S\,{\sc ii}]$\lambda$6717/H$\alpha$). H\,{\sc ii} regions ($diamonds$) show a perfect correlation, but the DIG locations ($asterisk$), are scattered; 
the opposite is observed in spiral galaxies.    }
         \label{fig7}
   \end{figure}

\begin{figure}
   \centering
   \includegraphics[width=15cm]{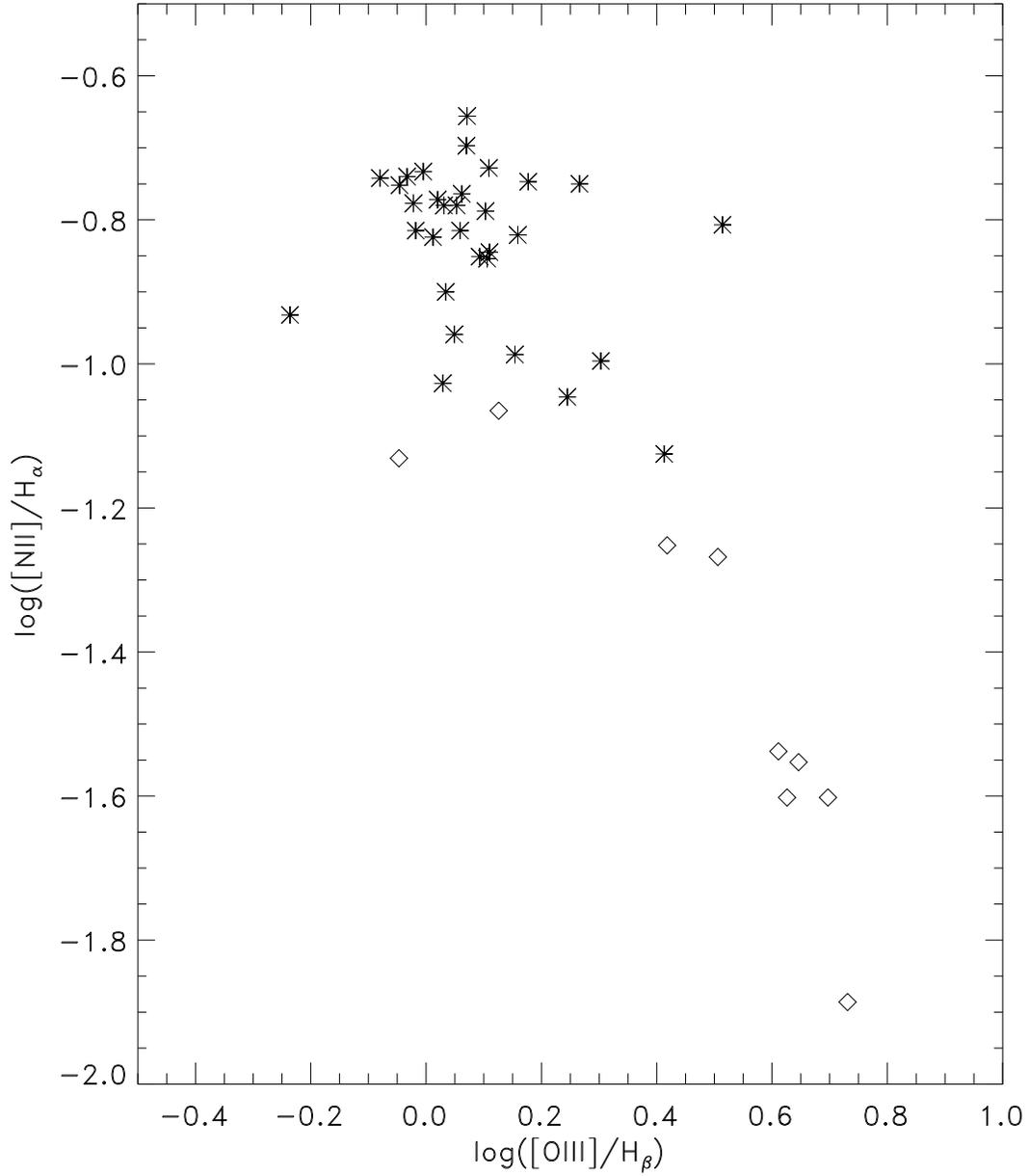}
      \caption{ log([O\,{\sc iii}]/H$\beta$) vs. log([N\,{\sc ii}]$\lambda$6584\AA/H$\alpha$) 
for all the data points. Symbols are as in Fig. ~\ref{fig7}. The H\,{\sc ii} 
regions follow a correlation, but the DIG locations are scattered,
with the majority of them between $-0.1$ and $0.3$ in log[O\,{\sc iii}]/H$\beta$. }
         \label{fig8}
   \end{figure}

Although classical photoionization models cannot explain the line ratios observed in NGC 6822, the correlation between [N\,{\sc ii}] versus [S\,{\sc ii}] and the recombination rates does not allow us to rule out these models completely. The strong correlation observed in the DIG of spiral galaxies between [N\,{\sc ii}] $\lambda$6583 and [S\,{\sc ii}] $\lambda$6717 indicates that the  [N\,{\sc ii}]/[S\,{\sc ii}] ratio is constant  (e.g. Otte \& Dettmar 1999; Rand 1998). Such a constancy cannot be explained by classical photoionization models (Hoopes \& Walterbos 2003). Using this result, Otte et al. (2001, 2002) derived an increase of temperature toward the halo that can be reproduced with the  multidimensional photoionization simulations of the density-bounded H\,{\sc ii} regions of Wood \& Mathis (2004). They concluded that the increase 
with distance from the plane observed in spiral galaxies is due to  
hardening of the radiation field with {\it z}. Such a correlation is not observed for any of the dwarf irregular galaxies studied so far (Hidalgo-G\'amez 2005; 2006; Hidalgo-G\'amez \& Flores-Fajardo 2007; A.M. Hidalgo-G\'amez 2007, in preparation). In Figure  ~\ref{fig7} this relationship is shown for this galaxy. A strong correlation is obtained for the H\,{\sc ii} regions ($diamonds$), with a regression coefficient of $0.99$ but only nine data points, while DIG locations ($asterisk$) show a vertical line. The explanations could be several: one might be related to the fact that  the locations with larger [N\,{\sc ii}] $\lambda$6584/H$\alpha$ also  might have larger excitation. Therefore, and since the ionization parameter of S$^{++}$ is smaller than that of O$^{++}$, most of the S$^+$ has been ionized into S$^{++}$, which is the reason for the behaviour of the DIG data points in Figure ~\ref{fig7} (M. Peimbert 2005, private communication). Figure ~\ref{fig8} shows the relation between log([N\,{\sc ii}] $\lambda$6584/H$\alpha$) and log([O\,{\sc iii}] $\lambda$5007/H$\beta$). A strong correlation for the DIG data point is expected if the above reasoning explains the lack of correlation. On the contrary, there is a stronger correlation for H\,{\sc ii} regions, while only a weak trend is observed for the DIG data points. Other explanations have to be invoked: a difference in the chemical content of any of these ions, or effects due to the 
different ionization level of these two species (Petuchowski \& Bennet 1993, 1995), 
may account for the variations observed in Figure ~\ref{fig7}. However, if the correlation 
is due to $T_e$, as in Wood \& Mathis (2004), Figure ~\ref{fig7} indicates that the 
temperature of the DIG is quite homogeneous, while significant differences are found 
in the H\,{\sc ii} regions. Figure 8 in HGOM01 shows that such differences
in the temperature are present inside these H\,{\sc ii} regions, especially inside Hubble V.

The second disconcerting factor is the high recombination rate determined 
from the He\,{\sc i} $\lambda$5876 line, following the expression (e.g. Miller 
\& Veilleux 2003) $$ {\Xi_{He}\over\Xi_{H}} = { 0.052 ({N_p \over N_{He^+}}) 
({T_e \over 8000})^{0.14} \over 0.048 ({He/H \over 0.1})}$$ where $\Xi_{He}$ and $\Xi_{H}$ are the emissivities of helium and hydrogen, respectively, $T_e$ is 
the electronic temperature, $He/H$ is the helium abundance, and 
$0.052 N_p/N_{He^+}$ comes from the emissivities of the hydrogen and helium 
recombination lines (Osterbrock 1989). The He abundances have been 
determined as He$^+$/H$\alpha$ because, as mentioned before, H$\beta$ 
might suffer from severe absorption. The electronic temperature is from 
Table 1. Recombination rates at the DIG locations between $0.65$ and $1$ are found, depending 
on the density of protons relative to He$^+$ and $T_e$, indicating that photoionization is important there.  

In recent years, leaky models have been proposed to explain the line 
ratios observed at the DIG. Hoppes \& Walterbos (2003) and Wood \& Mathis 
(2004) used density-bounded H\,{\sc ii} regions in their models, as did Domg\"orgen \& 
Mathis (1994). In this case, there is not enough nearby neutral gas to absorb the total Lyman continuum of the source and form a classical Str\"omgren sphere.  Wood \& Mathis (2004) 
focused on explaining the constant [N\,{\sc ii}]/[S\,{\sc ii}] ratio 
(not observed in NGC 6822). They also used the [O\,{\sc iii}]/H$\alpha$ ratio, which is very much affected by extinction. Moreover, although the [O\,{\sc iii}]/H$\alpha$ 
increases by a factor of $2$ in their models of density-bounded regions 
at low metallicities, as shown in their Figure 12, the  values of their models for the other line ratios are not low enough to fit our results listed in Table 1. Therefore, such model is not used  hereafter. In addition, the values obtained from the Domg\"orgen \& Mathis (1994) model  do not fit any of the line ratios obtained for NGC 6822.

The values of the [O\,{\sc iii}]/H$\beta$, [N\,{\sc ii}]/H$\alpha$ and [S\,{\sc ii}]/H$\alpha$ ratios of the DIG from Table 1 are compared with the leakage models from Hoopes \& Walterbos (2003). We make use of their Figure 14, in which they calculate the line ratios only for Orion metallicity and $log~ q = -3$, where $q$ is related to the ionization parameter. The He\,{\sc i}/H$\alpha$ ratio is not used, as our values lies outside the reach of the models. Moreover, several facts have to be remembered: due to the smaller metallicity of NGC 6822, some corrections to these models have to be carried out. The effect of lower metallicities increases the value of {\it q}, which in turn implies larger excitation and ionization. Moreover, irregular galaxies normally show a deficiency 
in nitrogen as compared to spirals (Hidalgo-G\'amez \& Olofsson 2002). As 
a consequence, the models predict a higher [N\,{\sc ii}] $\lambda$6584/H$\alpha$ line ratio than that observed in NGC 6822. Finally, 
a leakage of photons produces a harder spectrum and, therefore, lager 
values of the [O\,{\sc iii}]/H$\beta$ (and [O\,{\sc ii}]/H$\beta$) at lower $T_{ion}$.  In addition, and because of the smaller S/N of DIG E, the values of DIG W and C, 
if they differ, are preferred.  
 
As the excitation is very similar for both DIG W and C, a value of $1.2$ can be 
used. For  such a value,  a leakage from $45$\% at $38,000$ K to $60$\% at 
$42,000$ K is given by the model. The excitation of DIG E is much higher, 
with values of $41,000$-$50,000$ K for the same leakage range. 
The ratio [S\,{\sc ii}] $\lambda\lambda$(6717+6731)/H$\alpha$ is almost identical 
for DIG W and C: $0.12$. For a T$_{ion}$ of $40,000$ K, the 
leakage is $30$\%, increasing up $45$\% for $30,000$ K. 
Although DIG E has a lower ratio, there is no difference in the 
numbers. Something similar happens for the 
[N\,{\sc ii}] $\lambda$6584/H$\alpha$ ratio, where differences in the ratio 
do not imply differences in $T_{ion}$. For the three regions, a leakage of 
only $30$\% at $30,000$ K is obtained. Considering the arguments given above, 
a leakage of $40$\%-$50$\% with temperatures of $35,000$ K can fit both 
the excitation and the [S\,{\sc ii}]/H$\alpha$. Very similar values of $40$\%-$50$\% 
of photon leakage were obtained by Rela\~no et al. (2002) for 
NGC 346, the most luminous H\,{\sc ii} region in the Small Magellanic Cloud. The main problem is that 
nitrogen cannot be fitted. Leakages of $40\%$-$50\%$ give [N\,{\sc ii}]/H$\alpha$ 
ratios of $0.4$-$0.5$, which are not observed at any location. An explanation 
could be the low nitrogen content in irregular galaxies.

Another important caveat is that the leakage might have no preferable 
direction, and therefore the region in between the two H\,{\sc ii} regions 
might receive a large number of photons. Therefore, larger values of the excitation, [N\,{\sc ii}] $\lambda$6584/H$\alpha$ and [S\,{\sc ii}] $\lambda$6717/H$\alpha$, could be expected, which is not the case. On the contrary, there are no significant differences between the various DIG locations, either for the one-dimensional spectra or for the integrated ones. One might think that differences in the density will vary the leakage in different 
directions. As discussed above, we cannot say much about variations in the 
density. If it were larger in DIG C, then the leakage of photons would be less likely in  that direction, leading to similar line ratios 
everywhere. 

Finally, we should be concerned about some discrepancy between the values of the integrated spectra and the spectra along the slit. The main reason for the use of the former is the higher S/N ratio. It is well known that when the S/N is lower than $\approx 5$, the intensity of the line is overestimated (Rola \& Pelat 1994). An S/N lower than $5$ is typical for the one-dimensional spectra, and therefore we think that the line ratios of the integrated spectra are more reliable. In any case, we can do a double-check. First, we can obtain an average value from the one-dimensional spectra for each of the line ratios of interest. Both [N\,{\sc ii}]/H$\alpha$ and [S\,{\sc ii}]/H$\alpha$ are identical, and differences of $0.2$ compared to the values listed in Table 1 are found in the [O\,{\sc iii}]/H$\beta$. There is 
no difference in the ionization temperature or the photon-leaking percentage from the Hoopes \& Walterbos models. 

The second approach is to obtain a range of values for $T_{ion}$ and photon leaking for each of the one-dimensional spectra. Again, there are no differences compared to the results from the integrated spectra. Percentages between $40$\% and $60$\%, and values of T$_{ion}$ between $38,000$ and $41,000$ K, are obtained for most of the DIG locations from the [O\,{\sc iii}]/H$\beta$ ratio. Only three spectra ($10 \%$) have ionization temperatures larger than $41,000$ K. Values of $30\%$-$40\%$ of photons leakage and of $30,000$ to $36,000$ K for T$_{ion}$ are obtained from the [S\,{\sc ii}]/H$\alpha$ ratio for all the one-dimensional spectra. Therefore,it can be concluded that the leakages and T$_{ion}$ values found from the integrated spectra are correct for $90 \%$ of the DIG locations. 

In any case, according to us a leaky model is the model that most successfully 
explains the line ratios observed in the DIG in NGC 6822. The values of the leakage obtained here are similar to those of other irregular galaxies (Hidalgo-G\'amez 2005; 2006). 

\subsection{Other source of ionization}

The only discordant point to the picture presented in 4.1  is the low value of the [N\,{\sc ii}]/H$\alpha$ ratio. In order to fit it,  
an ionization source that radiates its energy at low ionization species is needed. 
The ideal model is the turbulence produced by the mixing of layers of 
different temperatures (Slavin et al. 1993). A model with depleted 
abundances, an intermediate temperature after mixing of $100,000$K, and a 
cloud temperature of $T_c = 10^4$ K will give values of [S\,{\sc ii}]/H$\alpha$ 
of $0.15$-$0.17$ and [N\,{\sc ii}] $\lambda$6584/H$\alpha$ of $0.09$-$0.11$ 
for velocities of hot gas between $20$ and $100$ km s$^{-1}$. These values 
are close to the values from the integrated spectra. The main problem is that 
the cold gas lines are broad with a non-Gaussian shape, something that was not observed. 

Finally, it might be interesting to explore the existence of shock waves 
that might have some influence on the ionization. According to de 
Blok \& Walter (2000), the H\,{\sc i} appearance of NGC 6822 is very disturbed. 
There are two main features: a giant H\,{\sc i} hole to the southeast and a cloud 
to the northwest, which might be  interacting with the galaxy and possibly  very important sources of shock waves. Even if both of them are far 
away from the region under study here, the disturbance of the ISM  over 
the entire galaxy can amplify them.

The best line for studying shocks is [O\,{\sc i}]$\lambda$6300\AA. Unfortunately, this line was not detected in any of the DIG locations, and therefore the [S\,{\sc ii}]/H$\alpha$ ratio must be used. This ratio has served for detecting shocks, e.g. bubbles around W-R stars (Polcaro et al. 1995). We can compare the values 
of this ratio with those from a shock model (Dopita \& Sutherland 1995). According to them, the [S\,{\sc ii}] $\lambda$6717/H$\alpha$ ratio varies between $0.26$ and $0.59$ for shocks with velocities from $150$ to $500$ km s$^{-1}$. When precursors are included in the model, those values become $0.18$ and $0.33$, respectively. As the  observed ratio is lower than $0.15$ ($0.17$ for non-reddening correction)  at any location along the slit, it can be concluded that the disturbed ISM of 
this galaxy has very little impact on the ionization of the DIG at the 
position of our observations.

\begin{figure}
   \centering
   \includegraphics[width=15cm]{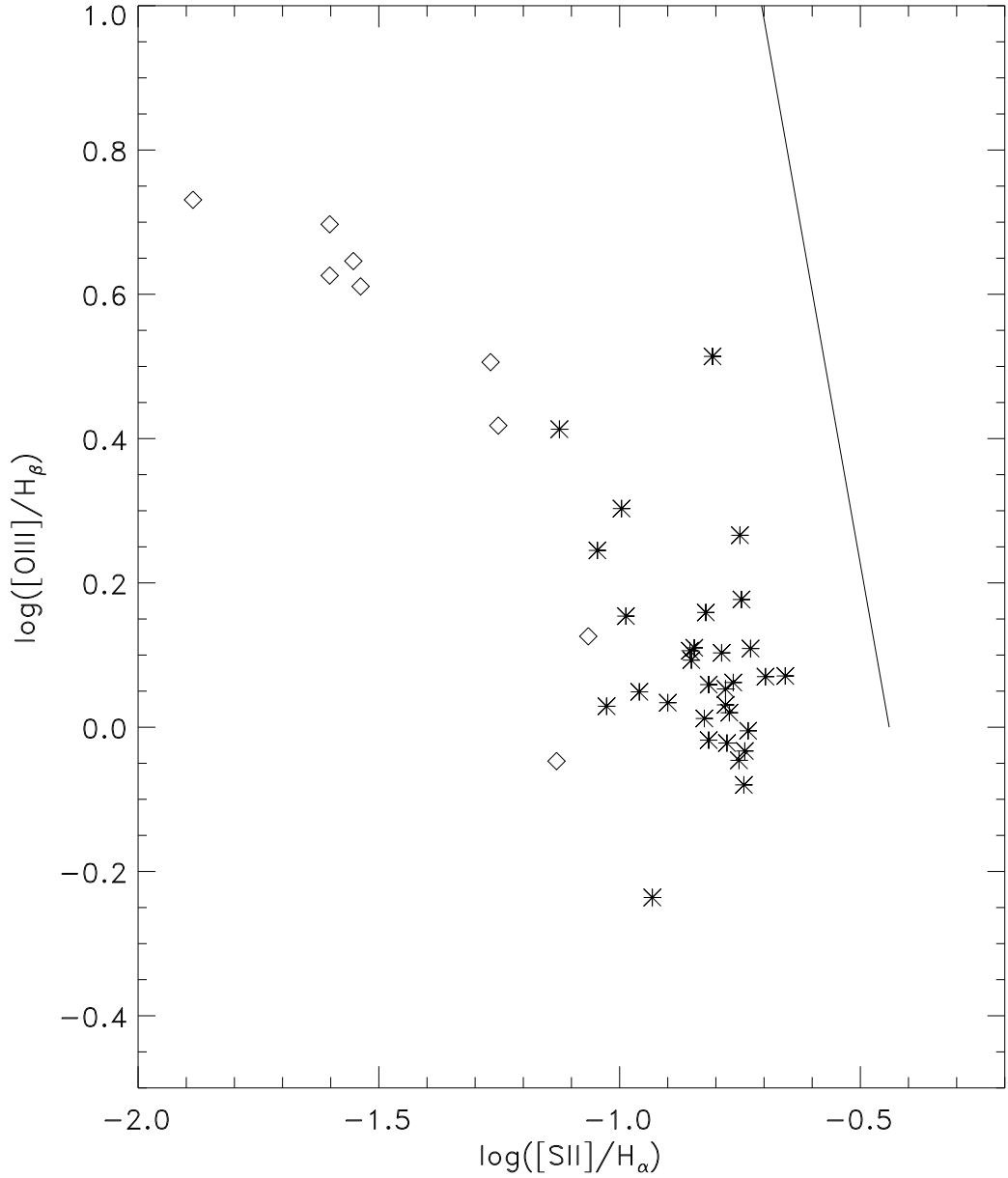}
      \caption{log([S\,{\sc ii}]$\lambda\lambda$(6717+6731)/H$\alpha$) vs. 
log([O\,{\sc iii}]/H$\beta$). The line corresponds to the shock model, without 
precursors, of Dopita \& Sutherland (1995). Symbols are as in Fig. ~\ref{fig7}. 
None of the DIG or the H\,{\sc ii} locations is situated in the shocked 
region ($to the right of the line$).  }
         \label{fig9}
   \end{figure}

Another useful tool for determining the influence of shocks could be the diagnostic diagram log [S\,{\sc ii}] $\lambda\lambda$(6717+6731)/H$\alpha$ versus 
log [O\,{\sc iii}]/H$\beta$. According to Veilleux \& Osterbrock (1987), this is the 
best diagram to differentiate shocked and photoionized regions. In Figure ~\ref{fig9} this diagram is shown for the data points considered in 
this investigation. In addition, the track of Dopita \& Sutherland's (1995) model without precursors and corrected for the metallicity of NGC 6822 is also
shown. It divides the plot into the shocked region ($to the right of the line$) and 
the non-shocked one ($to the left of the line$). All the data points for both H\,{\sc ii} 
and DIG lie inside the non-shocked region. Therefore, it can be concluded 
that shocks are not the dominant ionization source of the DIG in 
NGC 6822. 

\section{Discussion}

In this section we discuss the geometry of the DIG. The most likely 
dispostion is a homogeneous disposition of the ionized gas in the region. Like in Orion (O'Dell 2001), it is also probable that the gas thickness in front of the H\,{\sc ii} region is much smaller than that at the back. The excitation and the density can be used to discriminate between these possibilities. 

Unfortunately, there is much information that cannot be obtained from the [S\,{\sc ii}]$\lambda$6717/[S\,{\sc ii}]$\lambda$6731 ratio due to  the problems mentioned above (see Section 3). The excitation will show narrow peaks for the thin scenario. The thinner the layer, the narrower are the peaks. The main reason is that for a thin front layer, the contribution of the classical photoionization in the H\,{\sc ii} regions and their immediate vicinity is observed, while for a thick distribution it becomes diluted, making the photon-leakage ionization more 
important.  Also, this scenario will produce smaller C$_{\beta}$ values at the DIG locations. The main problem is that the C$_{\beta}$ values can also contain a contribution from the underlying stellar absorption. It becomes more important as the EM gets smaller and also due to the intensity of H$\beta$. Consequently, a severe stellar absorption will cause a higher extinction at the DIG locations, just as  a thick and homogenous distribution of the ionized gas will appear. 

Figure ~\ref{fig4} clearly shows that the peaks are narrow and do not extend very far from the H\,{\sc ii} regions. Therefore, no significant  quantity of gas should be located in front of the 
regions, in order to dilute the excitation.  In any case, in order to know 
the real geometry of the gas, more observations are needed, especially at 
radio and IR wavelengths. They will discriminate between the 
stellar absorption and a thick layer of gas in front of the regions. 
   
\section{Conclusions}

In the present investigation we have studied the characteristics of the DIG 
in the northern part of the irregular galaxy NGC 6822 using long-slit spectroscopy from FORS1 at the VLT. In order to study the properties of this gas, first we have to distinguish between those places that form part of the 
H\,{\sc ii} regions and those of the proper DIG. We used the cumulative function in the surface brightness in H$\alpha$. This function shows a change in the slope that marks the transition between H\,{\sc ii} regions and DIG locations. The value obtained, log~SB(H$\alpha$) = -$15.3$, is larger than those obtained for other irregular galaxies studied, such as IC 10 (Hidalgo-G\'amez 2005),  but similar to the value for DDO 50 (A.M. Hidalgo-G\'amez 2007, in preparation). 

We have studied the excitation, [N\,{\sc ii}] $\lambda$6584/H$\alpha$, 
[S\,{\sc ii}] $\lambda$6717/H$\alpha$, and [S\,{\sc ii}] $\lambda$6717/[S\,{\sc ii}]$\lambda$6731 along the 
slit. NGC 6822 is a very typical irregular galaxy as far as the DIG 
properties are concerned. Neither [N\,{\sc ii}]/H$\alpha$ nor [S\,{\sc ii}]/H$\alpha$ has values higher 
than $0.3$ at any location, while they are larger than $0.4$ 
in half of the spiral galaxies (see Table 1 in Hidalgo-G\'amez 2004). The
excitation is very high. In fact, only three locations have [O\,{\sc iii}]/H$\beta$ 
smaller than $1$; in contrast, it is very small ($\approx 0.1$) in those spirals 
where it has been measured (Galarza et al. 1999). Moreover, there is a weak 
correlation between [N\,{\sc ii}]/H$\alpha$ and [S\,{\sc ii}]/H$\alpha$ for the DIG locations.

Additionally, these ratios are useful for determining the ionization source(s) of the DIG. The line ratios are very different from 
the classical photoionization models (e.g. Domg\"orgen \& Mathis 1994), 
but helium is completely ionized and the [N\,{\sc ii}]/H$\alpha$ is too low. Therefore, the picture that might be more likely is the following: the helium is ionized 
in the vicinity of the H\,{\sc ii} regions while turbulence can increase the 
[N\,{\sc ii}]/H$\alpha$ ratio. Both the [O\,{\sc iii}]/H$\beta$ and [S\,{\sc ii}]/H$\alpha$ are increased by  photons escaping from the H\,{\sc ii} regions, which amounts to $40\%$-$50$\%.  Although  the H\,{\sc i} in NGC 6822 is very disturbed (de Blok \& Walter 2000), the disturbances have no consequence for the ionization of the DIG at the locations of Hubble X and Hubble V. 

Finally, it should be kept in mind that we are observing the result of all 
the processes simultaneously, and probably through a thick layer of neutral 
gas.

\begin{acknowledgements}

The authors are indebted to M. Peimbert and M. T. Ruiz for help with the 
observations and many fruitful discussions. A. M. H.-G. thanks M. Peimbert, 
who suggested including these data on the study of the DIG, and to many 
colleagues at Instituto de Astronom\'{\i}a-Universidad Nacional Aut\'onoma de M\'exico for fruitful comments on this work.   This investigation 
was partly supported by CONACyT project 2002-c40366 and DGAPA-UNAM grant 
IN 118405. An anonymous referee is thanked for interesting comments that have improved this manuscript.

\end{acknowledgements}

\clearpage

\clearpage

\clearpage

\clearpage

\clearpage

\clearpage

\clearpage

\clearpage

\clearpage

\clearpage

\begin{table*}
\scriptsize
\caption[]{ Total values of the line ratios (extinction corrected) from the integrated spectra. } 
\vspace{0.05cm}
\begin{center}
\begin{tabular}{c c c c c  c}
\hline
{Parameter }   & {DIG W} & {Hubble V}  & {DIG C} & {Hubble X} & {DIG E}  \\ 
\hline 

$\lbrack$OIII$\rbrack$$ \lambda$5007/H$\beta$ & 1.34$\pm$0.07 & 5.41$\pm$0.3 & 1.16$\pm$0.06 & 4.10$\pm$0.2 & 2.332$\pm$0.1 \\
He\,{\sc i} $\lambda$5875/H$\alpha$ & 0.07$\pm$0.03 & 0.036$\pm$0.003  & 0.06$\pm$0.03  & 0.040$\pm$0.003   & 0.06$\pm$0.03 \\
$\lbrack$NII$\rbrack$ $\lambda$6583/H$\alpha$ & 0.14$\pm$0.01 & 0.017$\pm$0.001 & 0.14$\pm$0.005 & 0.031$\pm$0.001   & 0.109$\pm$0.009  \\
$\lbrack$SII$\rbrack$ $\lambda$6717/H$\alpha$ & 0.123$\pm$0.006 & 0.028$\pm$0.001 & 0.120$\pm$0.006 & 0.045$\pm$0.002 & 0.112$\pm$0.006  \\
$\lbrack$SII$\rbrack$ $\lambda$6717/$\lbrack$SII$\rbrack$$\lambda$6731\AA & 1.44$\pm$0.09 & 1.6$\pm$0.07 & 1.4$\pm$0.07 & 1.41$\pm$0.07   & 1.829$\pm$0.09 \\ 
$\lbrack$NII$\rbrack$ $\lambda$6584/$\lbrack$SII$\rbrack$$\lambda$6717\AA & 1.14$\pm$0.1 & 0.62$\pm$0.05 & 1.17$\pm$0.06 & 0.69$\pm$0.05  & 0.97$\pm$0.09\\
\hline
$C_{\beta}$ & 2.75$\pm$0.01  & 0.44$\pm$0.005 & 2.71$\pm$0.01 & 0.08$\pm$0.005  & 0.708$\pm$0.001  \\
T$_e$ DIG (N) & 9228   &   -    & 8578    &   -   & 9936   \\
T$_e$ DIG (O) & 10202  &   -    & 9261    &   -   & 9698  \\
S/N (H$\alpha$) & 8 & 23 & 9 & 21& 6 \\
\hline
\end{tabular}
\end{center}
\end{table*}


\begin{thebibliography}{}

\bibitem[2000]{deblok} de Blok, W.J.G. \& Walter, F. 2000, ApJ, 537, 95

\bibitem[1971]{brocklehurst} Brocklehurst, M. 1971, MNRAS, 153, 471

\bibitem[2004]{castellanos} Castellanos, M., Valls-Gabaud, D., Diaz, A.I., \& 
Tenorio-Tagle, G. 2004, ``How does the Galaxy work?'' eds: E.J. Alfaro, E. Perez 
\& J. Franco, Kluwer Academic Publishers, p.101    

\bibitem[1994]{domgorgen} Domg\"orgen, H. \& Mathis. J.S. 1994, ApJ, 428, 647

\bibitem[1993]{dopitaa} Dopita, M.A. 1993. PASAu, 10, 359

\bibitem[1995]{dopitab} Dopita, M. A. \& Sutherland, R. 1995, ApJ, 455, 468

\bibitem[2005]{elwert} Elwert, T. \& Dettmar, R.-J. 2005, ApJ, 632, 277

\bibitem[199]{galarza} Galarza, V.C., Walterbos, R.A.M., \& Braun, R. 1999, AJ, 118, 2775
 
\bibitem[1998]{geenawalt} Greenawalt, B., Walterbos, R.A.M., Thilker, D. \& Hoopes, C.G. 1998, ApJ, 506, 135

\bibitem[2004]{hidalgo-gamezf} Hidalgo-G\'amez, A.M. 2004, ``How does the Galaxy work?'' eds: E.J. Alfaro, E. Perez \& J. Franco, Kluwer Academic Publishers, p.71    

\bibitem[2005]{hidalgo-gameza} Hidalgo-G\'amez, A.M. 2005 A\&A, 442, 443 

\bibitem[2005]{hidalgo-gamezb} Hidalgo-G\'amez, A.M. 2006, AJ, 131, 2078

\bibitem[1998]{hidalgo-gamezc} Hidalgo-G\'amez, A.M., \& Olofsson, K. 1998, A\&A, 334, 45

\bibitem[2001]{hidalgo-gamezd} Hidalgo-G\'amez, A.M., Olofsson, K. \& Masegosa, J., 2001, A\&A, 367, 388 (HGMO)

\bibitem[2002]{hidalgo-gameze} Hidalgo-G\'amez, A.M. \& Olofsson, K. 2002, A\&A, 389, 836 

\bibitem[2006]{hidalgo-gamezg} Hidalgo-G\'amez, A.M. \& Ram\'{\i}rez-Fuentes, D., A\&A submitted

\bibitem[2007]{hidalgo-gamezh} Hidalgo-G\'amez, A.M. \& Flores-Fajardo, N. 2007, AJ submitted

\bibitem[1989]{hodge} Hodge, P., Lee, M.G., \& Kennicutt, R.C.Jr. 1989, PASP, 101, 32

\bibitem[2003]{hoopes} Hoppes, C. G. \& Walterbos, R.A.M. 2003, ApJ, 586, 902

\bibitem[1990]{hunter} Hunter, D.A. \& Gallagher, J.S. 1990, ApJ, 362, 480

\bibitem[2004]{hunter} Hunter, D. A., \& Elmegreen, B.G. 2004, AJ, 128, 2170

\bibitem[2005]{karachentsev}  Karachentsev, I.D. 2005, AJ, 129, 178

\bibitem[1997]{martin} Martin, C.L. 1997, ApJ, 491, 561

\bibitem[1986]{mathis}Mathis, J.S. 1986, ApJ, 301, 423

\bibitem[1985]{mcall} McCall, M.L., Rybski, P.M., \& Shields, G.A. 1985, ApJS, 57, 1

\bibitem[2003]{miller} Miller, S.T. \& Veilleux, S. 2003, ApJ, 592, 79

\bibitem[2001]{o'dell} O'Dell, C.R. 2001, ARA\&A, 39, 99

\bibitem[1989]{osterbrock} Osterbrock, D.E. 1989, Astrophysics of Gaseous 
 Nebulae and Active Galactic Nuclei, (University Science Books, Mill Valley, 
CA) 
\bibitem[1999]{otte} Otte, B. \& Dettmar, R-J. 1999, A\&A, 343, 705

\bibitem[2001]{ottea} Otte, B., Reynolds, R.J., Gallagher, J.S. III, \& Ferguson, A.M.N. 2001, ApJ, 560, 207

\bibitem[2002]{otteb} Otte, B., Gallagher, J.S.III, \& Reynolds, R.J. 2002, ApJ, 572, 823

\bibitem[2005]{peimbert} Peimbert, A., Peimbert, M. \& Ruiz, M.T. 2005, ApJ, 634,1056 

\bibitem[1993]{petuchowski93} Petuchowski, S.J. \& Bennett, C.L. 1993, ApJ, 405, 591

\bibitem[1995]{petuchowski95} Petuchowski, S.J. \& Bennett, C.L. 1995, ApJ, 438, 735

\bibitem[Polcaro et a. (1995)]{pol95} Polcaro, V.F., Rossi, C., Norci, L., \& Viotti, R. 1995, A\&A, 303, 211

\bibitem[1998] {rand} Rand, R.J. 1998, ApJ, 501, 137

\bibitem[2002]{relano} Rela\~no, M., Peimbert, M., \& Beckman, J. 2002, ApJ, 564, 704

\bibitem[1984]{reynolds} Reynolds, R.J. 1984, ApJ, 282, 191

\bibitem[1999]{reynoldsb} Reynolds, R.J., Haffner, L.M.,\& Tufte, S.L. 1999, ApJL, 525, 21

\bibitem[1994]{rola} Rola, C. \& Pelat, D. 1994, A\&A, 287, 676

\bibitem[1979]{savage} Savage, B.D. \& Mathis, J.S. 1979, ARA\&A, 17, 73

\bibitem[1993]{slavina} Slavin, J.D, Shull, J.M. \& Begelman, M.C. 1993, ApJ, 407, 83

\bibitem[1939]{stromgren} Str\"omgren, B. 1939, ApJ, 89, 526
 
\bibitem[2000]{tullman} T\"ullman, R., \& Dettmar, R.-J. 2000, A\&A, 362, 119

\bibitem[1991]{RC3} de Vaucouleurs, G., de Vaucouleurs, A., Corwin, H.G., et al. 1991, Third Reference Catalogue of Bright Galaxies (Springer-Verlag)

\bibitem[1989]{veilleux} Veilleux, S. \& Osterbrock, D.E. 1987, ApJS, 63, 295

\bibitem[2001]{wyder} Wyder, T.K. 2001, AJ, 122, 2490

\bibitem[2004]{wood} Wood, K., \& Mathis, J.S. 2004, MNRAS, 353, 1126
\end{thebibliography}
\end{document}